\documentclass[twocolumn,showpacs,preprintnumbers,amsmath,amssymb]{revtex4}
\usepackage{graphicx}
 \newcommand\la{\langle}
 \newcommand\ra{\rangle}
 \newcommand\beq{\begin{equation}}
 
 \newcommand\eeq{\end{equation}}
 \newcommand\beqn{\begin{eqnarray}}
 \newcommand\eeqn{\end{eqnarray}}
 \newcommand\GeV{{\rm GeV}}
 
\def\GeV{\,\mbox{GeV}}
\def\TeV{\,\mbox{TeV}}

\def\lsim{\mathrel{\rlap{\lower4pt\hbox{\hskip1pt$\sim$}}
    \raise1pt\hbox{$<$}}}         %less than or approx. symbol
\def\gsim{\mathrel{\rlap{\lower4pt\hbox{\hskip1pt$\sim$}}
    \raise1pt\hbox{$>$}}}         %greater than or approx. symbol
\def\BA{\begin{eqnarray}}
\def\BE{\begin{equation}}
\def\BF{\begin{figure}[htb]}
\def\BT{\begin{table}[htb]}
\def\EA{\end{eqnarray}}
\def\EE{\end{equation}}
\def\EF{\end{figure}}
\def\ET{\end{table}}

\def\la{\langle}
\def\ra{\rangle}

\begin{document}

\title{Heavy quarkonium  in saturated environment of high-multiplicity pp collisions}
%\title{Influence of saturation on heavy quarkonium production\\ in high-multiplicity pp collisions}

\author{B. Z. Kopeliovich$^1$}
\author{H. J. Pirner$^{2}$}
\author{ I. K. Potashnikova$^1$}
\author{K. Reygers$^{3}$}
\author{Iv\'an Schmidt$^{1}$}

\affiliation{\centerline{$^1$Departamento de F\'{\i}sica,
Universidad T\'ecnica Federico Santa Mar\'{\i}a; }
%{Centro Cient\'ifico-Tecnol\'ogico de Valpara\'iso;
{Avenida Espa\~na 1680, Valpara\'iso, Chile}
\\
{$^{2}$Institute for Theoretical Physics, University of Heidelberg, Germany
}\\
{$^3$Physikalisches Institut, University of Heidelberg, Germany}}

\begin{abstract}

High-multiplicity $pp$ collisions exhibit features,  traditionally  associated with nuclear effects. Coherence motivates to treat high-multiplicity $pp$, $pA$ and $AA$ collisions on an equal footing. 
We rely on the phenomenological parametrization for mean multiplicities of light hadrons and $J/\psi$, assuming their linear dependence on $N_{coll}$ in $pA$ collisions. 
The results of this approach underestimate the recently measured production rate of $J/\psi$ at very high hadronic multiplicities.
The linear dependence of $J/\psi$ multiplicity on $N_{coll}$ is subject to  predicted nonlinear corrections, related to mutual boosting of the saturation scales in colliding dense parton clouds. A parameter-free calculation of the non-linear corrections allows to explain data for $p_T$-integrated yield of $J/\psi$ at high hadronic multiplicities.
Calculations are in a good accord with data binned in several $p_T$-intervals as well.
As was predicted, $\Upsilon$ and $J/\psi$ are equally suppressed at forward rapidities in $pA$ collisions. Consequently, their fractional multiplicities at forward rapidities in  $pp$ collisions are equal as well, and their magnitude  agrees with data.

\end{abstract}

%\date{\today}


\maketitle

\section{Introduction}

The popular models of multiparticle production in $pp$ and $pA$ collisions are based on the eikonal multi-Pomeron exchange, or Glauber models. On the contrary to the wide spread believe, the Glauber model contains no clue to the multiplicity distribution. This is a model for the elastic scattering amplitude, which is related to the total cross section by the unitarity relation. One can also calculate the total inelastic cross section, however no relation between the multiple scattering terms in the elastic amplitude and inelastic processes follows from the Glauber model.
The unitarity relations between each term in the Glauber elastic amplitude, expanded over multiple interactions, and the corresponding inelastic processes,
was proposed by Abramovsky, Gribov and Kancheli  \cite{agk}, known as AGK cutting rules.

The unitarity cut of the elastic scattering amplitude can be done simultaneously through several Pomerons, while the other uncut Pomerons play role of absorptive corrections. 
Those corrections, also known as shadowing, lead to a substantial reduction of the total inelastic cross section. However,  they cancel each other and do not affect the inclusive particle production cross section. This peculiar feature of absorption corrections to the inclusive cross section is known as AGK cancellation \cite{agk}. 

Relating different unitarity cuts of the elastic amplitude with multiplicity of produced particles (any species) one finds that multiplicity is proportional to the number of cut Pomerons, usually called number of collisions, $N_{coll}$. 
Therefore, the normalised multiplicities
of light hadrons and of $J/\psi$, defined respectively as,
\beqn
R_{h}\equiv \frac{dN_{h}/dy}{\left\la dN_{h}/dy\right\ra},
\ \ \ \ 
R_{J/\Psi}&\equiv& \frac{dN_{J/\Psi}/dy}{\left\la dN_{J/\Psi}/dy\right\ra},
\label{200}
\eeqn
 are expected to be equal, $R_{J/\Psi}=R_{h}$, and proportional to $N_{coll}$.

This expectation is in apparent contradiction with data \cite{alice-psi-1,mult-new,anton1,anton2}, presented below in Fig.~\ref{fig:data}, demonstrating a significantly steeper rise of $R_{J/\Psi}$ with hadron multiplicity in comparison with $R_{h}$. Moreover, it has been known since long time ago \cite{huefner} that the simple relation $R_h=N_{coll}$, offered by the  eikonal model, is strongly broken, and the multiplicity dependence on $N_{coll}$ for light hadrons produced in $pA$ collisions, is usually parametrized as,
\beq
R_h=1+\beta_h(N_\mathrm{coll}-1),
\label{300}
\eeq
with $\beta_h<1$ fitted to data. In the Glauber model $N_{coll}$ is given by
\beq
N_{coll}=A\,\frac{\sigma_{in}^{pp}}{\sigma_{in}^{pA}}
\label{315}
\eeq

The analysis of available data on $pA$ collisions performed in \cite{huefner} demonstrated consistency between different fits and no evidence for energy dependence of $\beta_h$.
The multiplicity measured  at $\sqrt{s}=5\TeV$ in \cite{alice-mult}, $dN_h^{pA}/dy=17.24\pm 0.66$, which leads to $\beta_h\approx0.55$. This is the value we rely upon in what follows.

The  breakdown of the AGK cancellation is caused by coherence effects in gluon radiation \cite{kst1}.
The AGK rules \cite{agk} assume the Bethe-Heitler regime of particle production, which e.g. leads to
$n$-times higher rapidity density of multiplicity $dN_h/dy$ for $n$ cut Pomerons compared to
a single cut. This regime is known to be broken by the Landau-Pomeranchuk effect of coherent radiation from multiple interactions. The reduction of gluon radiation caused by coherence is also known as gluon shadowing \cite{kst2}. Only the accumulated mean transverse momentum
of the radiated gluons keeps memory of multiple interactions, which lead to $p_T$-broadening of radiated gluons, usually called saturation or color-glass condensate \cite{mv}. 

Besides broken  $N_{coll}$-dependence, Eq.~(\ref{300}), coherent production of quark-antiquark pairs also leads to a strong deviations from  universality of $R_h$.
At first glance this might look surprising, because gluons are expected to hadronize at long distances, where no final state interaction is possible.
Such a simplified interpretation of space time development is not correct.

As an example, $\bar qq$ production in DIS at small $x$ exhibits  shadowing due to propagation and attenuation of the $\bar qq$ dipole through
the whole target \cite{krt-shad}. This shadowing is a higher twist, so it is weaker for heavy than for light quarks.

Production of $\bar qq$ by gluons is more involved, both the incoming gluon and produced $\bar qq$ interact with the target and screen each other.
As a result, shadowing of quark production also is a higher twist and is described as propagation of an artificial 3-body dipole $|\bar qqg\ra$. Correspondingly, charm quarks \cite{kt-hf}, charmonia \cite{nontrivial,rhic-lhc,marat} are shadowed considerably less than light quarks. This explains in particular the observed non-universality  $R_{J/\Psi}>R_{h}$.

The effects of coherence possess a strong scale dependence for the relation between fractional 
multiplicity and number of collisions. If the popular  parametrization
Eq.~(\ref{300}) is applied to $J/\psi$,
\beq
R_{J/\psi}=1+\beta_{J/\psi}(N_\mathrm{coll}-1),
\label{400}
\eeq
one should expect a larger value of the parameter $\beta_{J/\psi}$.

Thus, from (\ref{300})-(\ref{400}) we get,
\beq
\frac{R_{J/\Psi}-1}{R_h-1}=\frac{\beta_{J/\psi}}{\beta_h}
\label{380}
\eeq
The difference between the parameters $\beta_{J/\psi}>\beta_h$, leads to a steeper rise of 
$R_{J/\Psi}$ in comparison with $R_h$, as is demonstrated by data.

The upper bound for the speed of rise for the ratio (\ref{380}) is
\beq
\frac{R_{J/\Psi}-1}{R_h-1} \leq 
\frac{1}{\beta_h} \approx 1.8.
\label{385}
\eeq
The processes of least shadowing corrections, like production of Drell-Yan pairs, $\Upsilon$, $B$ and $D$ mesons, are expected to approach this bound.

We do not attempt here at theoretical evaluation of $\beta_{J/\psi}$ (though is doable \cite{marat}), but prefer to rely on a fit to data. 
Another popular parametrization applied to measured $A$-dependence of $J/\psi$ production is, 
\beq
R^A_{J/\psi}=N_{coll}\,A^{\alpha-1},
\label{500}
\eeq
where the fitted parameter $\alpha$ turns out to be  close to unity, implying a  weak nuclear suppression of $J/\psi$. Expanding this expression in small parameter $1-\alpha\ll1$
we can relate $\beta_{J/\psi}$ in (\ref{400}) and $\alpha$,
\beq
1-\beta_{J/\psi}=(1-\alpha)\,\ln A\,\frac{N_\mathrm{coll}}{N_\mathrm{coll}-1}
\approx (1-\alpha)\,\ln A,
\eeq
assuming a large number of collisions.

The exponent $\alpha$ in (\ref{500}) also does not expose a strong energy dependence.
It was accurately measured at $\alpha=0.95$ in the fixed target experiment E866 at $\sqrt{s}=40\GeV$  \cite{e866}. This value agrees with $\alpha= 0.9 - 0.98$ measured at the mid-rapidity at $\sqrt{s}=5.02\TeV$ \cite{alice3,alice3a}, or interpolated between forward and backward rapidities \cite{alice1,alice2,alice4}.

Thus, according to (\ref{380}) $R_{J/\Psi}$ rises linearly with increasing $R_h$, or nearly linearly if to rely on Eq.~(\ref{300}), as was done in \cite{kpprs}. The dependence of 
$R_{J/\Psi}$ on $R_h$ shown in Fig.~\ref{fig:data} by dashed curves was calculated at $\sqrt{s}=5.02\TeV$, with $\beta_h=0.55$ and
the interval values of $\alpha=0.95-0.98$, which we consider as the corridor of the current uncertainty in measured $A$-dependence of $J/\psi$ production. 

%%%%%%%%%%%%%%%%
\begin{figure}[htb]
\vspace{-10mm}
\begin{center}
  \includegraphics[width=8cm]{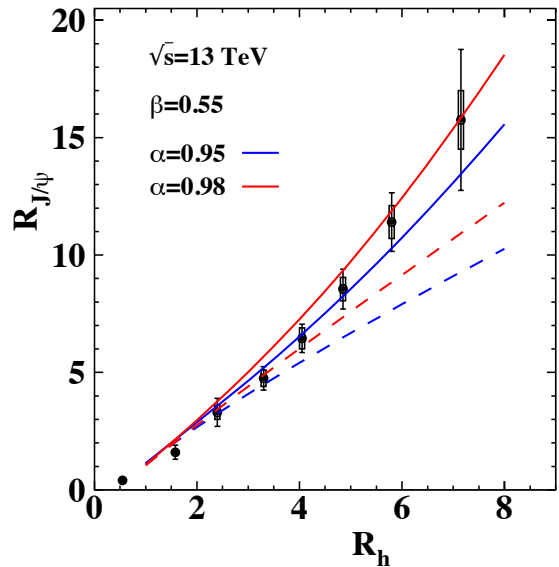}   \\[-15mm]
 \caption{\label{fig:data} (Color online) Normalized multiplicity of $J/\Psi$, $R_{J/\Psi}$, vs normalized multiplicity of charged hadrons, $R_h$. Data are {\sl preliminary} from ALICE \cite{mult-new,anton1,anton2} for $pp$ collisions at $\sqrt{s}=13\TeV$ at for $y<0.9$.
 The dashed curves are calculated with Eqs.~(\ref{380}), (\ref{500}), while solid curve incorporate the effect of mutually boosted saturation scale. In each pair the upper and bottom curves are calculated with $\alpha=0.98$ and $0.95$ respectively.
 }
  \end{center}
 \end{figure}
%%%%%%%%%%%%%%%%

While these results agree with data at $R_h\leq 4$, available at the time of publication of \cite{kpprs}, new measurements at $\sqrt{s}=13\TeV$ and higher multiplicity up to $R_h=7$ \cite{mult-new,anton1,anton2} considerably exceed the predicted values of $R_{J/\Psi}$, as one can see in Fig.~\ref{fig:data}. As we discussed above, neither $\beta_{J/\psi}$, nor
$\beta_h$ demonstate any significant energy dependence, therefore comparison of results in Fig.~\ref{fig:data} at different energies is legitimate.

\section{Effects of gluon saturation}

Multiple interactions naturally lead to broadening of the transverse momentum distribution,
which has been calculated  in \cite{jkt,saturation} for proton-nucleus collisions in agreement with data, and predicted for high-multiplicity $pp$ collisions in \cite{kpprs}. In the leading order broadening is proportional to the number of collisions with known coefficient,
\beq
\Delta p_T^2 = \frac{9\,C(E)}{2\,\sigma_\mathrm{in}^{pp}}\,(N_\mathrm{coll}-1),
\label{780}
\eeq
where $N_\mathrm{coll}$ is related to $R_h$ by Eq.~(\ref{400}).

The energy dependent factor $C(E)$ in (\ref{780}) controls the cross section $\sigma_{\bar qq}(r)$of interaction of a small size $r$ colorless dipole of energy $E$ (in the nuclear rest frame) \cite{jkt}
\beq
C(E)={1\over2}\,\vec\nabla_{r_1}\!\!\!\cdot\vec\nabla_{r_2}\,\sigma_{\bar qq}(\vec r_1-\vec r_2,E)
\Biggr|_{\vec r_1=\vec r_2}.
\label{740}
\eeq
Notice that for broadening of gluons, which
are responsible for $J/\psi$ production,
the pQCD factor $9/4$ is introduced in (\ref{780}).

Since parton model interpretation of high-energy processes is not Lorentz invariant (only observable are), broadening of the $p_T$-distribution, which looks like a result of multiple interactions in the nuclear rest frame, is interpreted as saturation of small-$x$ parton density in the nucleus in its infinite momentum frame. Moreover the value of the saturation scale is directly related to the magnitude of the broadening, $Q_s^2=\Delta p_T^2$ \cite{saturation}.

Apparently, the additional kicks gained by the parton from multiple collisions, increase the effective scale of the process $Q^2\Rightarrow Q^2+\Delta p_T^2$ \cite{boosting,puzzles,nontrivial,rhic-lhc,marat}. As a result of DGLAP evolution the nuclear gluon density $g_A(x,Q^2+\Delta p_T^2)$
turns out to be suppressed at large $x$, but enhanced at small $x$ \cite{boosting}. 

Thus, the rate of $J/\psi$ production in high-multiplicity $pp$ collisions turns out to to be enhanced 
by gluon saturation and rises with $R_h$ steeper in comparison with Eq.~(\ref{380}).

In $pA$ collisions only the projectile proton undergoes multiple interactions, which modify its parton distribution function (PDF), 
while the PDFs of bound nucleons remain unchanged. In the case of $AA$, or high multiplicity 
$pp$ collisions, the interaction becomes symmetric, both assembles of colliding constituents are subject to multiple interactions, increasing their partonic content at small $x_{1,2}$.
On the other hand, multiple interactions in a denser  gluonic medium become more intensive, leading to further increase of gluon density. Thus, a rising small-$x$ gluon density in one of the colliding protons, induced by multiple interactions at large $R_h$, stimulates stronger broadening of the partons in another colliding proton, which in turn leads to further increase of gluon density in the first proton.
Such a mutual boosting of the gluon densities and saturation scales in the colliding protons at large $R_h$ is similar to the rise of the saturation scales in colliding nuclei is described in more detail in \cite{boosting}, and satisfies the bootstrap equation,
\beq
\tilde Q_s^2 = \frac{3\pi^2}{2}\,\alpha_s(\tilde Q_s^2+Q_0^2)\,
xg(x,\tilde Q_s^2+Q_0^2)\, 
\frac{N_{coll}}{\sigma_{in}^{pp}},
\label{750}
\eeq
where $N_{coll}$ is related to $R_h$ by (\ref{300}). The characteristic scale $Q_0$ is the merging  line between the nonperturbative and perturbative regimes of broadening. While the  the latter is described perturbatively, the former is not calculable, but relies  on the dipole phenomenology \cite{saturation}.
\beq
\frac{3\pi^2}{2}\,\alpha_s(Q_0^2)\,
xg(x,Q_0^2) = C(E),
\label{755}
\eeq
where $C(E)$ is given by Eq.~(\ref{740}) and $E=Q_0^2/2m_N x$.
Such a strategy is similar to what is used in DGLAP based analyses of data, except 
our evolution equation (\ref{750}) is essentially nonlinear.

\begin{figure}[htb]
\vspace{-10mm}
\begin{center}
  \includegraphics[width=8cm]{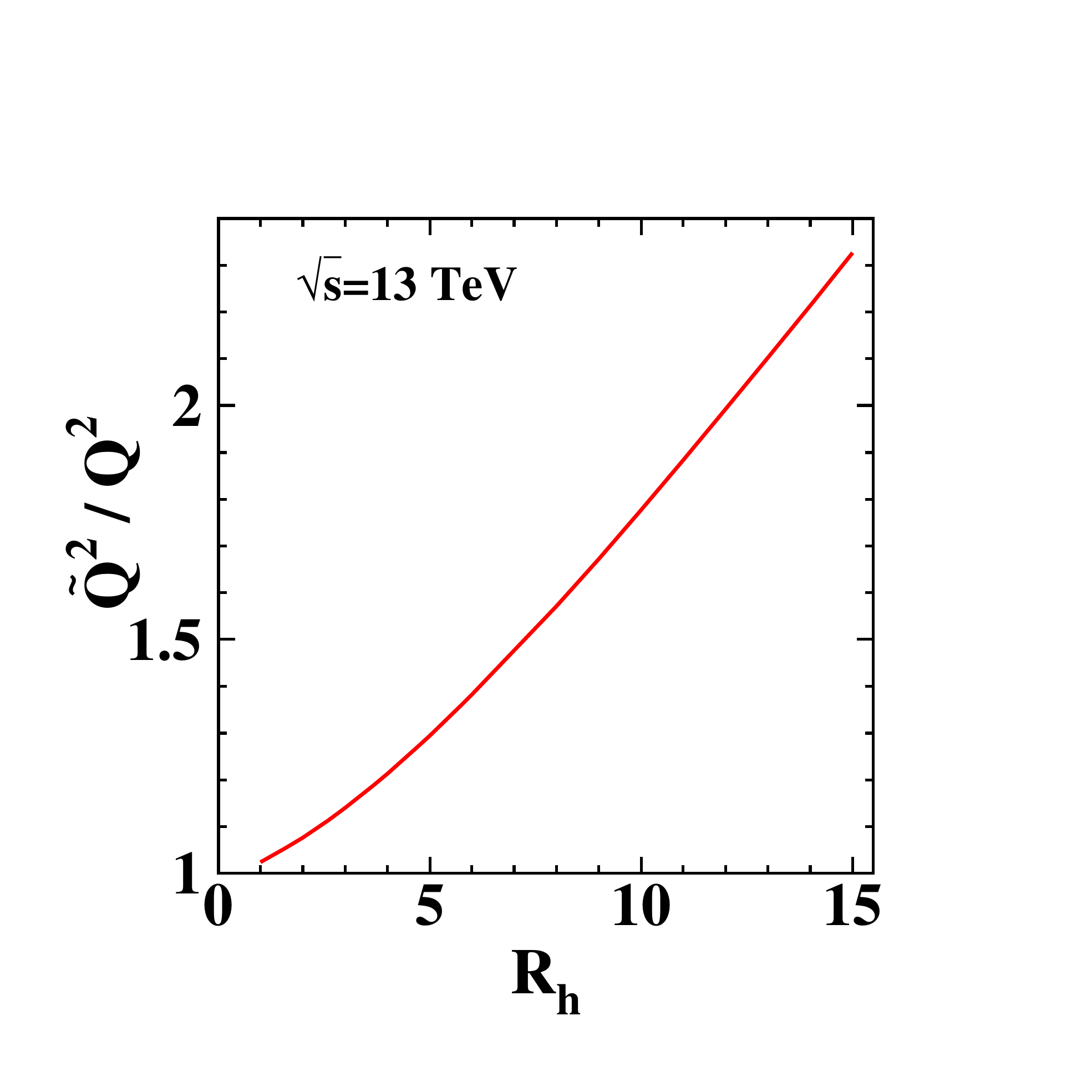}   
  \\[-5mm]
 \caption{\label{fig:Q-tilde} (Color online) Boosted saturation scale Eq.~(\ref{750}) in the proton relative to the original scale $Q^2=4m_c^2$ vs fractional multiplicity of light hadrons.
 }
  \end{center}
 \end{figure}

The saturated scale calculated with Eq.~(\ref{750}) for  high-multiplicity $pp$ collisions relative to the characteristic scale
$Q^2=4m_c^2$ vs $R_h$ is shown in Fig.~\ref{fig:Q-tilde}.
%%%%%%%%%%%%%%%%
%%%%%%%%%%%%%%%%
At high multiplicities the boosted scale significantly exceeds the original scale, 
generating via DGLAP evolution more gluons at small $x$ and a larger yield of $J/\psi$.

Notice that increase of the saturation scale hardly affects the hadron multiplicity
$R_h$, because hadrons with low $p_T$, which give the main contribution to the multiplicity, are already produced with parton densities, which are at the unitarity bound. This is why $\beta_h$ show
no evidence for energy dependence.

%\section{\boldmath$J/\psi$ production in a saturated environment}

\section{Transverse momentum distribution}

An increase of the scale of the process at small $x$ leads to a steep growth of the gluon density and as a result to enhanced production of $J/\psi$. Therefore the relative production rate $R_{J/\psi}$ plotted in Fig.~\ref{fig:data} by dashed curves, should be corrected for the effects of saturation multiplying by the factor $g_N(x,\tilde Q_s^2+Q^2)/g_N(x,Q^2)$.
The results are depicted in Fig.~\ref{fig:data} by solid curves.
Apparently, introduction of the effects of saturation improved agreement with data at high multiplicities.

It is also instructive to compare our results and data in different intervals of  transverse momentum $p_T$ os $J/\psi$. 
\beqn
R_{J/\psi}(p_T) = \frac{\int_{p_T^{min}}^{p_T^{max}} dp_T p_T\,
dN^{pp}_{J/\psi}(R_h)/dy d^2p_T}
{\int_{p_T^{min}}^{p_T^{max}} dp_T p_T
\left\la
dN^{pp}_{J/\psi}/dy d^2p_T
\right\ra}
\label{655}
\eeqn
Here the multiplicity of $J/\psi$ production in the nominator is taken at a given fractional multiplicity $R_h$ of light hadrons, while in the denominator the $J/\psi$ production rate
is summed over hadron multiplicity. To perform integrations in (\ref{655}) one needs to know the explicit form of $p_T$-dependence of the cross section.
We rely on the popular parametrization,
\beq
\frac{dN^{pp}_{J/\psi}}{dy d^2p_T}=
\frac{dN^{pp}_{J/\psi}}{dy}
\frac{1}{\la p_T^2\ra}
\left(1+\frac{p_T^2}{(n-2)\la p_T^2\ra}\right)^{-n}
\label{665}
\eeq
The mean transverse momentum squared can be extracted directly from data \cite{alice2}
applying a simple interpolation procedure. For $\sqrt{s}=13\TeV$ we arrived at $\la p_T^2\ra=11.72\GeV^2$.
Fixing this value and the shape (\ref{665}) we fitted the remaining parameter at $n=3.2$.

We apply the same parametrization of the $p_T$-dependence to the numerator of
(\ref{655}), but increase the mean value of $p_T^2$ by broadening,
$\la p_T^2\ra \Rightarrow  \la p_T^2\ra+\Delta p_T^2$, which was calculated with
Eq.~(\ref{780}).

\begin{figure}[htb]
\vspace{-10mm}
\begin{center}
 \includegraphics[width=8cm]{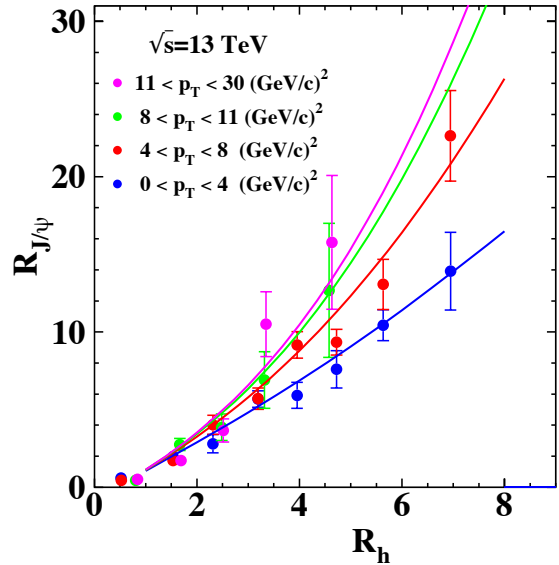} \\[-15mm]
 \caption{\label{fig:data-pt-new} (Color online) Fractional multiplicity $R_{J/\Psi}$, vs fractional multiplicity of charged hadrons, $R_h$, for different intervals of $J/\psi$ transverse momentum $p_T$ indicated in the plot. Data are {\sl preliminary} from \cite{mult-new,anton1,anton2}, the curves are calculated with Eq.~(\ref{655})
including the effects of saturation and mutual boosting of the saturation scale. }
  \end{center}
 \end{figure}

The data for $R_{J/\psi}$ measured in different $p_T$ intervals are depicted in Fig.~\ref{fig:data-pt-new} together with results of our calculations of Eq.~(\ref{655})
including the effects of saturation and mutual boosting of the saturation scale.
%%%%%%%%%%%%%%%%
%%%%%%%%%%%%%%%%

\section{\boldmath$J/\psi$ and $\Upsilon$ at forward rapidities}

Data show that $J/\psi$ production rate in $pA$ collisions is more suppressed  at forward than at the mid-rapidity \cite{e866,alice1,alice4}. For the interval $2.5<y<4$, measured in \cite{alice1}, the suppression factor $A^{J/\psi}_{eff}/A\approx 0.65$, which corresponds to $\alpha=0.92$.

Although the radius of $\Upsilon$ is smaller than of $J/\psi$, and $\Upsilon$ is less suppressed at the mid-rapidity, the $\bar bb$ dipole cross section
rises  with energy faster than $\bar cc$ and $A^{\Upsilon}_{eff}$ is  steeply falling towards forward rapidities (see Fig.~4 in \cite{nontrivial}). At  rapidities $2.5<y<4$ higher-twist shadowing is predicted \cite{nontrivial} to suppress $\Upsilon$ as much as $J/\psi$, $A^{\Upsilon}_{eff}\approx A^{J/\psi}_{eff}= 0.65A$.

Notice that  the parameters $\beta_{J/\psi}=\beta_{\Upsilon}=0.57$, are only slightly above the hadronic value $\beta_h=0.55$, so the originally expected relation $R_{J/\psi}=R_h$ looks well satisfied. Of course, this does not mean restoration of the eikonal model, this is just a numerical coincidence.

\begin{figure}[htb]
\vspace{-0mm}
\begin{center}
 \includegraphics[width=8cm]{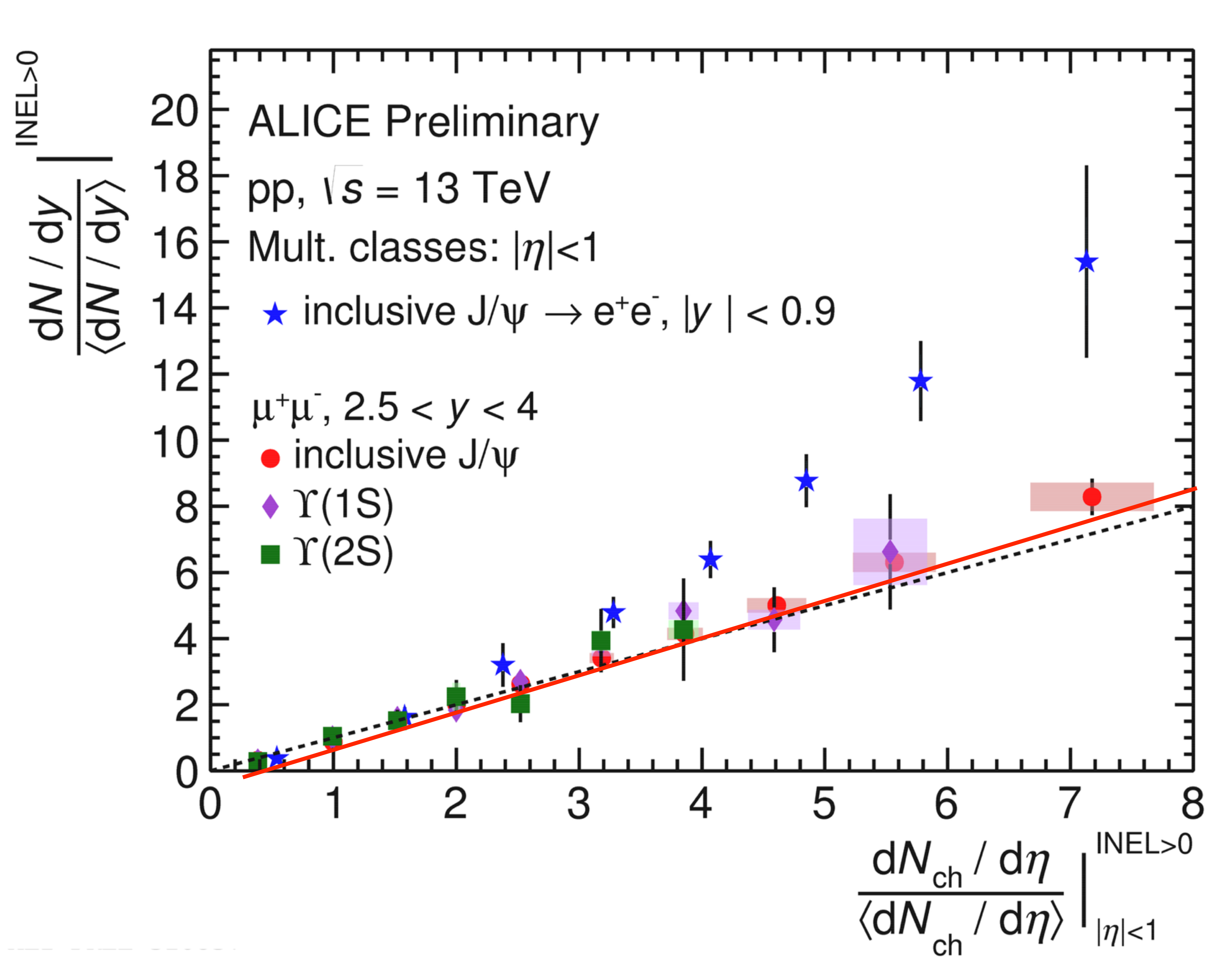} \\
 %[-15mm]
 \caption{\label{fig:forward} (Color online) Fractional multiplicities $R_{J/\Psi}$ and $R_{\Upsilon}$ at $2.5<y<4$, vs fractional multiplicity of charged hadrons, $R_h$ at $|\eta|<0.9$. 
  Data are {\sl preliminary} from \cite{mult-new,anton1,anton2} at $\sqrt{s}=13\TeV$. The solid red curve is calculated with Eq.~(\ref{380}) with $\beta_{J/\psi}=\beta_{\Upsilon}=0.47$.}
  \end{center}
 \end{figure}

Applying  relation (\ref{380}) we get $R_{J/\psi}=R_{\Upsilon}$ vs $R_h$ depicted by red solid curve in Fig.~\ref{fig:forward}. Notice that it does not need to be corrected for the boosted saturation effects, because the correction is small at these rapidities (see Fig.~6 in \cite{nontrivial}).
%%%%
%%%%%

\section{Summary}

High-multiplicity $pp$ collisions exhibit features, which traditionally are associated with nuclear effects. Motivated by a long coherence time of interaction, we observe a close similarity between multiple interactions in $pp$, $pA$ and $AA$ collisions. In order to enhance multiple interactions in the former case one should trigger on high multiplicity of produced hadrons, while nuclei allow to reach the same multiplicity easier, by increasing the number of collisions Eq.~(\ref{315}).  We employ the phenomenological description of the mean multiplicity in $pA$ collisions, Eq.~(\ref{300}), violating the AGK cutting rules. Fitted to data,
the observed nuclear effects for $J/\Psi$ production, enable one to predict the multiplicity dependence of the $J/\Psi$ production rate in $pp$ collisions. However,
the linear relation Eq.~(\ref{380}) between the fractional multiplicities $R_{J/\psi}$ and $R_h$ underestimate the yield of $J/\psi$ at large hadron multiplicities
\cite{mult-new,anton1,anton2}. 

Nevertheless, the linear dependence of the $J/\psi$ production rate on $N_{coll}$ is subject to nonlinear corrections, 
predicted in \cite{boosting}. They are related to mutual increase of the saturation scales in a colliding nuclei \cite{boosting}, or two dense parton clouds.
The parameter-free calculation of the nonlinear corrections allows to explain the observed nonlinearity
of $R_{J/\psi}$ vs $R_h$, as well as data binned in several $p_T$-intervals.

A sensitive test of the present model can be made with recently measured \cite{mult-new,anton1,anton2} yield of $J/\psi$ at forward rapidities, which demonstrate similar fractional multiplicities $R_{J/\psi} = R_h$. 
We explain this result within our phenomenological description, while it might be a challenge for other available models. Besides, we predicted the same relation for heavy bottomia,
$R_{J/\psi} = R_{\Upsilon}$, which is also in a good accord with data \cite{mult-new,anton1,anton2}.

Notice that the correlation of multiplicities of $J/\psi$ vs hadrons observed at forward rapidities, are well described by Pythia \cite{pythia}. However, this is not a parameter-free approach, the parameters of the generator are adjusted to data to be explained.

%\vspace*{2cm}
\begin{acknowledgments}

\end{acknowledgments}
We are thankful to Anton Andronic and Sarah Porteboeuf-Houssais for informative discussions and providing links to released new data.
This work was supported in part by grants CONICYT - Chile FONDECYT 1170319 and 1180232, by USM-TH-342 grant, and by CONICYT - Chile PIA/BASAL FB0821.

\end{document}